%% file: paper.tex
\documentclass[letterpaper, conference]{ieeeconf}  

\IEEEoverridecommandlockouts                              
\usepackage{tablefootnote}
\usepackage{amssymb}
\usepackage{amsmath}
\usepackage{graphics}
\usepackage{graphicx}
\usepackage{epsfig}
\usepackage{epstopdf}
\usepackage{mathrsfs}
\usepackage{cite}
\usepackage[mathscr]{euscript}
\usepackage{nameref}

\include{philipMacros}

\newtheorem{theorem}{Theorem}[section]

\newtheorem{proposition}[theorem]{Proposition}
\newtheorem{lemma}[theorem]{Lemma}

\newtheorem{definition}{Definition}[section]

\begin{document}
%
\title{Designing for Emergent Security in Heterogeneous Human-Machine teams}

\author{{Philip N. Brown}
\thanks{P. N. Brown is an Assistant Professor with the Department of Computer Science at the University of Colorado, Colorado Springs, {\texttt{philip.brown@uccs.edu}. Corresponding author.}}
}


%


\maketitle

\begin{abstract}
This work seeks to design decisionmaking rules for autonomous agents to jointly influence and optimize the behavior of teamed human decisionmakers in the presence of an adversary.
We study a situation in which computational jobs are scheduled on servers by a collection of autonomous machines in concert with self-interested human decisionmakers, and the human and machine schedulers must react to an adversary's attack on one of the servers.
We show a simple machine scheduling policy such that if all schedulers have permission to schedule jobs on all servers, increasing the penetration of machine schedulers always increases the level of security in the system, even when the machine schedulers have no explicit coordination or communication amongst themselves.
However, we show a companion result in which simple constraints on server availability can nullify the machine schedulers’ ability to effectively influence human schedulers; here, even if machine schedulers control an overwhelming majority of jobs, are socially-aware, and fully coordinated amongst themselves, they are incapable of influencing human decisionmakers to mitigate the harm of an attack.
\end{abstract}


%
\IEEEpeerreviewmaketitle

\section{Introduction}

The recent rush of technology into society's daily life has promised to solve grand problems, but also may bring with it grand new risks.
New technological paradigms such as the Internet of Things allows people to interact with their devices in unprecedented ways, but increased means of interaction may increase vulnerability to adversarial manipulation~\cite{Khan2018,La2016}.
Increasingly, society and technology are participants in each other's affairs, which requires that engineers and computer scientists must be increasingly aware of the effects of each on the other.
Autonomous decisionmakers must be designed to interact well with human decisionmakers, and this may look fundamentally different from how autonomous decisionmakers typically interact with one another~\cite{Hamdi2014}.

In response to these new challenges, \emph{game theory} (the formal study of interactive decisionmaking) has emerged as a set of mathematical tools which promise to shed some light on the central design tradeoffs in this space~\cite{Pavel2012,Brown2017a,Pillai2016}.
In applying game theory to these types of systems, it is common to speak of ``emergent behavior'' in distributed decisionmaking; that is, game theory studies the aggregate behavior which \emph{emerges} from the entangled decisionmaking processes of autonomous machine agents, human users,  and strategic attackers~\cite{Laszka2014,An2017,Hota2018}.

This paper represents an initial study on a concept that we term \emph{emergent security}.
That is, we seek fundamental principles which will allow the principled design of autonomous decisionmakers in sociotechnical systems; these autonomous agents should react to the actions of attackers in an intelligent, strategically-aware way to mitigate the harm of attacks.

A strategic attacker's approach may be twofold: first, the attacker may \emph{directly} attack the system and reduce quality-of-service~\cite{Fawzi2014}.
Alternatively, the attacker may \emph{indirectly} reduce quality-of-service by influencing ordinary system users to engage in inefficient behavior; this can include attack methods such as manipulative ``social-engineering'' attacks~\cite{Brown2017f,Abraham2010}.

The defending autonomous agents' approach is similarly twofold: first, agents should be designed to directly avoid and mitigate the effects of the attacker's actions; second, the autonomous agents should be designed to influence the behavior of ordinary system users so that \emph{they} avoid and mitigate the effects of the attacker's actions~\cite{Ratliff2018,Brown2016d,Bolouki2017}.

This paper asks whether and under what circumstances heterogeneous teams of human and machine decisionmakers can act in concert to provide security for distributed systems, despite a lack of centralized coordination among the decisionmakers.
Specifically, we seek distributed decisionmaking architectures that provide \emph{emergent security}; 
that is, the distributed decisionmaking rules and interaction framework guarantee that aggregate emergent behavior in the system responds effectively to adversarial manipulation.

We pose a model of collaborative job scheduling on servers: there is a collection of $n$ servers and a constant inflow of jobs; each job requires scheduling on any of the servers.
Each job is given either to one of a collection of ``machine schedulers’’ (which then select a server for the job based on a pre-specified algorithm) or the job is self-scheduled (in which case the job is assigned to a server with minimum current service time).

We consider a situation in which an intelligent attacker, wishing to decrease system throughput, selects a server and directly degrades its performance (e.g., by performing a denial-of-service attack that increases average service time)~\cite{Amin2009}.
Subsequently, the machine schedulers and self-scheduled jobs may react to the increased service time on the attacked server and modify their scheduling choices accordingly.
The attacker’s hope is to maximize the resulting performance degradation, both \emph{directly} due to increased service times on the affected server and \emph{indirectly} by manipulating the human and machine scheduling policies to create a cascade of inefficiencies affecting other servers.

In light of this model, our question is simple: can the decision rules of the machine schedulers be designed in such a way that system security \emph{emerges} as the result of the interactive decisionmaking of the various schedulers?
We say that a system exhibits \emph{strong emergent security} if, for all attacks, the schedulers collectively respond optimally to the attack.
On the other hand, we say that a system exhibits \emph{weak emergent security} if, for all attacks, the performance of the schedulers' collective response is  at least no worse than a fixed scheduling policy.
Given these definitions, our results are threefold:

\vspace{2mm}
\noindent {\bf Symmetric scheduling:}
When every server is available to every scheduler and the machine schedulers apply a simple locally-optimal scheduling policy, every system exhibits strong emergent security provided that the fraction of jobs controlled by machine schedulers exceeds some threshold (where this threshold is strictly less than $1$).
Crucially, this threshold is independent of the \emph{number} of machine schedulers.

\vspace{2mm}
\noindent {\bf Symmetric scheduling and linear server delay:}
Here, if the $n$ servers each have identical linear delay functions, provided that at least $1/n$ of the jobs are controlled by machine schedulers, the collective response to any attack is optimal (i.e., the system has strong emergent security even if only  a small minority of jobs are machine-controlled).
Furthermore, every system with linear server delay functions exhibits weak emergent security. 
That is, regardless of the mass of jobs controlled by machine schedulers, we have that for any attack, collective scheduling behavior is always at least as good as a fixed scheduling policy that is unresponsive to attacks.

\vspace{2mm}
\noindent {\bf Constrained server availability:} 
However, the above results need not hold in the case that each scheduler can only access a subset of servers.
In this case, even if the majority of jobs are centrally controlled by a \emph{single} machine scheduler, the system may perform nearly as poorly as a completely unresponsive policy.
Here, though the system still exhibits weak emergent security, the ability of the machine schedulers to reduce system cost is severely curtailed.

Our results suggest that emergent security of heterogeneous human-machine teams is a valid goal and can be attained with simple decision rules.
However, designers must take care to understand how constraints and heterogeneity can subvert an otherwise performant system.

\section{Model}

We are given a set of $n$ servers and a total load of $n$ units of jobs requiring service.
Following from~\cite{Roughgarden2004}, each individual job is taken to be infinitesimally ``small,'' so that each job contributes a negligible amount to congestion.
We denote the \emph{load} (or mass of scheduled jobs) at server $i$ by $x_i\geq0$, and $\tau_i(x_i)$ denotes the delay suffered by a job at server $i$ under load $x_i$.
We adopt the usual convention that each $\tau_i(x_i)$ is convex, nondecreasing, and continuously differentiable, and assume that the uncongested service time at all servers is equal, or for all $i,j$, $\tau_i(0)=\tau_j(0)$.
A \emph{service profile} is denoted $x:=(x_1,x_2,\dots,x_n)$; we require that all jobs are serviced so that $\sum_ix_i=n$, which we denote by $x\in n\Delta(n)$%
\footnote{
Note that for $a>0$, we write $a\Delta(n)$ to denote the standard probability simplex with each of its constraints multiplied by $a$, so that  its vertices are of the form $(0,\dots,a\dots,0)$.
}.
Our key cost metric is average service delay, denoted\vs
\begin{equation}
\mathcal{C}(x) := \sum_{i=1}^n \frac{x_i\tau_i(x_i)}{n}.
\end{equation}

\subsection{Adversarial action}

An adversary, wishing to disrupt the system and increase system cost, selects a server (throughout this paper, we adopt the convention that this server is numbered $1$) and attacks it with strength $\alpha>0$.
The effect of this attack is to uniformly increase service times%
\footnote{Here, we consider an additive degradation of service.
Naturally, this degradation could take other forms.
We adopt this particular formulation for simplicity, postponing the study of other formulations for future work.}
 at server $1$ by an amount $\alpha$; this is modeled by replacing the nominal delay function $\tau_1(x_1)$ with the degraded delay function $\tau_1^\alpha(x_1) := \tau_1(x_1) + \alpha$.
For convenience, for any $i\geq1$, we write $\tau_i^\alpha(x_i) = \tau_i(x_i)$.

Since the attacker’s action directly increases service times for any job on server $1$, it induces a new system cost of\vs
\begin{align}
\mathcal{C}^\alpha(x) 	&:= \sum_{i=i}^n \frac{x_i\tau^\alpha_i(x_i)}{n} = \mathcal{C}(x) +  \frac{x_1\alpha}{n}. \label{eq:attackCost}
\end{align}

%
%
%

\subsection{Heterogeneous Human-Machine Teamed Scheduling} \label{ssec:team}

The main goal of this paper is to understand the effects of human-machine teaming on the security of distributed systems.
To accomplish this, we consider a case in which there are $m$ engineered schedulers (called \emph{machine} schedulers) acting in conjunction with a population of uncoordinated human schedulers.
The system designer endows the machine schedulers with decision rules with which they select a scheduling profile, whereas the human schedulers select a server merely on the basis of minimizing their own service delay.


Machine scheduler $k\in\{1,\dots,m\}$ is responsible for scheduling a mass of $r_k$ jobs, with $r:=\sum_{k=1}^mr_k\leq n$.
We refer to $r$ as the \emph{machine penetration level} of the system.
We term the remaining $n-r$ jobs \emph{selfish jobs,} and we model their ``behavior’’ by assuming that each infinitesimal job among the selfish jobs has a human owner that schedules it with the singular goal of minimizing its service time, given the attacker's action and the choices of other jobs. 

\begin{figure}
\centerline{\includegraphics[scale=0.25]{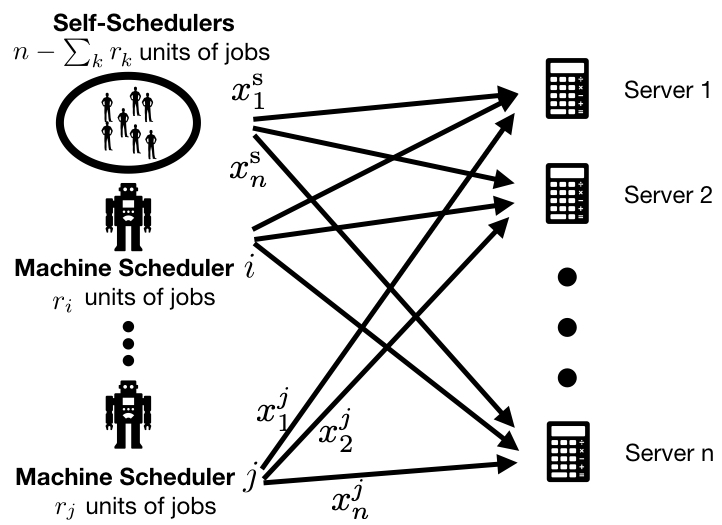}}
\caption{Depiction of teamed scheduling setup for the nominal model descried in Section~\ref{ssec:team} and considered in Theorems~\ref{thm:main} and~\ref{thm:special}.
Any of the $k$ machine schedulers and any of the self-schedulers are free to schedule a job on any of the $n$ servers.}
\label{fig:schem1}
\vspace{-5mm}
\end{figure}

Figure~\ref{fig:schem1} contains a simple depiction of the model.
We allow each machine scheduler to schedule on any of the $n$ servers, so each machine scheduler $k$ selects a service profile $x^k:=(x_1^k,\dots,x_n^k)\in x^k\in r_k\Delta(n)$, satisfying $x^k_i\geq0$ and $\sum_{i=1}^nx^k_i=r_k$.
Given a collection of machine-scheduled profiles $\{x^k\}$ and a selfishly-scheduled profile $x^{\rm s}$, the aggregate load on machine $i$ is given by $x_i= \sum_{k=1}^mx_i^k + x_i^{\rm s}$.
The aggregate scheduling profile resulting from teamed scheduling is represented $x=(x^{\rm s},x^1,\dots x^m)$.
Sometimes, to highlight the dependence of a scheduling profile on the actions of a particular machine $k$, we may also represent an aggregate profile as $x=(x^{\rm s},x^k, x^{-k})$, where $x^{-k}$ denotes the scheduling by machine schedulers other than $k$.



We model the aggregate behavior resulting from the collective decisionmaking of the $m$ machine schedulers and the selfish jobs by a concept that we term \emph{team equilibrium} $\xt$, or an aggregate scheduling profile satisfying
\begin{equation}
\xt^k  \in \arginf_{x^k\in r_k\Delta(n)} \C^\alpha\left(\xt^{\rm s},{x}^k,\xt^{-k}\right) \label{eq:optTeam}
\end{equation}
and 
\begin{align}
\xt^{\rm s}_i>0 &\implies \tau^\alpha_i(\xt_i) \leq \tau^\alpha_j(\xt_j) \ \forall j\in \{1,\dots,n\}. \label{eq:selfishTeam}
\end{align}
Note that at a team equilibrium,~\eqref{eq:optTeam} implies that each machine scheduler is scheduling (globally) optimally (given the choices of others), and~\eqref{eq:selfishTeam} implies that each selfish job is selecting a server with minimum delay (given the choices of others).
While equilibrium existence in arbitrary games is not guaranteed, known results for heterogeneous congestion games give us the following proposition:
\begin{proposition}
For any $\alpha\geq0$, any $n\geq2$, any $m\geq1$, and any $\{r^k\}_{k=1}^m$ satisfying $\sum_{k=1}^m r^k \leq n$, the above game has a team equilibrium $\xt$.
\end{proposition}

This follows immediately from elementary equilibrium existence results in nonatomic games~\cite{Harker1988,Orda1993,Mas-Colell1984}.

%

\subsection{Emergent Security}

We say that a system exhibits \emph{emergent security} if the teamed agents in the system (both human and engineered) naturally adapt to attacks in such a way that the system performance remains near-optimal even when the system is under attack.
Since the team equilibrium concept implicitly assumes no coordination between agents, it is a natural means to study the emergent security arising from various decision rules.

In this context, we distinguish between different degrees of security by comparing the performance of a team equilibrium against that of two benchmarks:

\subsubsection*{Optimal scheduling} Given $\alpha$, we denote an optimal scheduling policy by $x^*(\alpha)\in\argmin_x\C^\alpha(x)$.
The ultimate goal in the emergent security paradigm is for the performance of team equilibria to equal that of $x^*(\alpha)$.
\subsubsection*{Unresponsive scheduling} Here, we compare the performance of a team equilibrium to that of an unresponsive baseline scheduling policy that we denote $\xb(\alpha)$; this policy is optimal at no-attacks when $\alpha=0$, but is unresponsive to changes in $\alpha$.
That is, for all $\alpha\geq0$, $\xb(\alpha)=x^*(0)$.
Since this scheduling policy is constant in $\alpha$, we typically denote it simply $\xb$.

Using these two definitions, we can now state the main qualitative performance metric of the paper:

\begin{definition} \label{def:emsec}
We say that a system exhibits \emph{strong emergent security} if, for all $\alpha\geq0$, it holds that all team equilibria are optimal.
That is, if $\xt(\alpha)$ is a team equilibrium, we have that for all $\alpha\geq0$,
\begin{equation}
\C^\alpha(\xt(\alpha))=\C^\alpha(x^*(\alpha)). \label{eq:strong}
\end{equation}
On the other hand, we say that a system exhibits \emph{weak emergent security} if, for all $\alpha\geq0$, it holds that all team equilibria are no worse than unresponsive scheduling.
That is, if $\xt(\alpha)$ is a team equilibrium, we have that
\begin{equation}
\C^\alpha(\xt(\alpha))\leq \C^\alpha\left(\xb\right). \label{eq:weak}
\end{equation}
\end{definition}\vspace{2mm}

In each of our forthcoming contributions, we exhibit a system setting and then determine which of Definition~\ref{def:emsec}'s properties hold.

\section{Contributions} \label{sec:ourContributions}

\subsection{Team scheduling exhibits emergent security for symmetric schedulers}
%
%

Our first theorem is a general result on the effectiveness of heterogeneous human-machine teams; it establishes a baseline guarantee that performance guarantees for such systems are monotone nondecreasing in the penetration of machine schedulers.
Furthermore, for high-enough machine penetration, every system exhibits strong emergent security.
\begin{theorem} \label{thm:main}
For any $n\geq2$, $m\geq1$, $r\in[0,n]$, $(r^k)_{k=1}^m\in r\Delta(m)$ and any $\alpha\geq0$, let $\xt(\alpha)$ and $\xt’(\alpha)$ be team equilibria with machine penetration levels $r$ and $r’$, where $r>r’$. Then it holds that
\begin{equation}
\C^\alpha\left(\xt(\alpha)\right) \leq \C^\alpha\left(\xt’(\alpha)\right). \label{eq:generalineq}
\end{equation}
Furthermore, for every collection of delay functions $\{\tau_i\}$, there exists a machine penetration threshold $\bar{r}<n$ such that if $r\geq\bar{r}$, then the system exhibits strong emergent security.
That is, for all $\alpha\geq0$ the resulting team equilibrium (denoted $\xt(r,\alpha)$) is a globally optimal response: 
\begin{equation} 
\C^\alpha\left(\xt(r,\alpha)\right) = \inf_{x\in n\Delta(n)} \C^\alpha(x). \label{eq:globopt}
\end{equation}
\end{theorem}\vspace{2mm}
The proof of Theorem~\ref{thm:main} appears in the Appendix.\vspace{2mm}

That is, Theorem~\ref{thm:main} provides three things: first, when all schedulers have access to all servers, an implication of~\eqref{eq:generalineq} is that heterogeneous human-machine teams are at least as performant as self-scheduled jobs.
Second, for a fixed $\alpha$, increasing the penetration level of machine schedulers can never harm the performance of the system.
Third, regardless of the \emph{number} of machine schedulers, team scheduling is globally optimal provided that the fraction of self-scheduled jobs is small enough. 
Crucially here, every system may have a positive mass of self-scheduled jobs and still exhibit optimal team scheduling for all attack levels $\alpha$.

It is important to note that the results of Theorem~\ref{thm:main} do not rely in any way on explicit coordination or communication between the various machine schedulers.
The only coordination between machine schedulers happens implicitly through the machines' common objective function.

\subsection{Closed-form expressions for linear delay functions}

Theorem~\ref{thm:main} paints a broad picture of the benefits of machine scheduling (even if uncoordinated), but gives little hint as to the actual performance gains possible.
In particular, the only information given about the optimal machine penetration threshold $\bar{r}$ is that $\bar{r}<n$; but the question remains: how \emph{much less} than $n$?

Accordingly, our next theorem reports the effects of team scheduling in closed-form for a simplified setting.
Here, we consider the case that all servers have identical linear delay functions, or $\tau_i(x_i) = x_i$.
In this setting, it is simple to observe that the nominal optimal scheduling policy under no attack has $x_i=1$ for all $i$; since all delay functions are equal, the policy which equates the delay on every server is optimal.
The following theorem characterizes the effects of $n$, $r$ and $\alpha$ on the efficiency of team equilibria.

\begin{theorem} \label{thm:special}
For this system, given $\alpha$, it holds that 
\begin{align}
\bar{r} 		&= 1- \frac{\alpha(n-1)}{2n}  \nonumber 
			\leq 1.
\end{align}
When $r\geq\bar{r}$, the system exhibits strong emergent security and we have that every team equilibrium $\xt$ is \emph{globally optimal}, satisfying\vs\vs
\begin{equation}\label{eq:optC}
\C^\alpha(\xt) = \min\left\{\frac{n}{n-1},1 + \frac{\alpha}{n} - \frac{\alpha^2(n-1)}{4n^2}\right\}.
\end{equation}
When $r\in\left(\bar{r}-\frac{\alpha(n-1)}{2n},\bar{r}\right)$, we have that every associated team equilibrium $\xt$ strictly outperforms the fully-selfish equilibrium and satisfies
\begin{equation} \label{eq:strict}
\C^\alpha(\xt) = \min\left\{\frac{n}{n-1},\frac{r^2 + r\left(\frac{\alpha(n-1)}{n} -2\right) +n}{n-1}\right\}
\end{equation}
Finally, when $r\leq1-\frac{\alpha(n-1)}{n}$, we have that every associated team equilibrium $\xt$ is equal to a fully selfish equilibrium and satisfies\vs\vs
\begin{align}
\C^\alpha(\xt) 	&= \min\left\{ \frac{n}{n-1}, 1+\frac{\alpha}{n} \right\} \label{eq:selfish} \\
			&\leq \C^\alpha\left(\xb\right) = 1+\frac{\alpha}{n}. \label{eq:baselineSelf}
\end{align}
That is, for all $r\in[0,n]$, this system exhibits weak emergent security.

\end{theorem}\vspace{2mm}
The proof of Theorem~\ref{thm:special} appears in the Appendix.

\begin{figure}
\centerline{\includegraphics[scale=0.4]{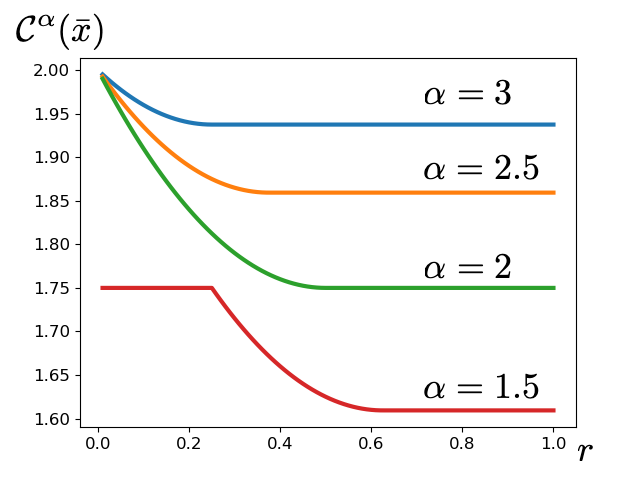}}
\caption{Normalized cost of a team equilibrium with $n=2$, plotted as a function of machine penetration $r$, for various values of $\alpha$.
Note that the cost is weakly decreasing in the level of machine penetration $r$, and the largest gains are made when $r$ is relatively low compared to the total mass of jobs in the system.}
\label{fig:CwrtR}
\vspace{-5mm}
\end{figure}

Several things are of note here.
First, the penetration threshold $\bar{r}$ is quite low: at most, only a $1/n$ fraction of jobs need to be under the control of machine schedulers to ensure that all team equilibria are optimal.
Second, no matter how many jobs are sent to machine schedulers, every team equilibrium is guaranteed to have a relative cost less than $n/(n-1)\leq 2$, regardless of the strength of the attack.
Compared to a baseline policy which is unresponsive to attacks, this represents a dramatic potential reduction in cost, since the unresponsive policy has a cost of $1+\alpha/n$, unbounded in $\alpha$.
Finally, equation~\eqref{eq:selfish} implies that if the machine scheduler penetration is too low, the selfish schedulers entirely dominate the scheduling profile and the machine schedulers have no effect.
In fact, for moderate attacks when $\alpha\leq n/(n-1)$,~\eqref{eq:baselineSelf} indicates that this situation is identical in cost to the unresponsive baseline scheduling policy~$\xb$ which simply ignores attacks.
Figure~\ref{fig:CwrtR} depicts the dependence of the normalized system cost on the machine penetration level $r$ for various attack levels $\alpha$.

\subsection{Loss of security due to scheduling constraints} \label{ssec:secLoss}

The result of Theorem~\ref{thm:main} is appealing in its simplicity: if all schedulers (both machine and selfish) have access to scheduling on all servers, then the optimal policy for each machine is simply to behave as though it were the only scheduler and schedule optimally given the choices of others.
In this section, we ask if the result would change if some servers were inaccessible to some of the schedulers; we show that even very simple scheduling constraints can render Theorem~\ref{thm:main}’s optimal scheduling policies almost entirely ineffective.

\begin{figure}
\centerline{\includegraphics[scale=0.23]{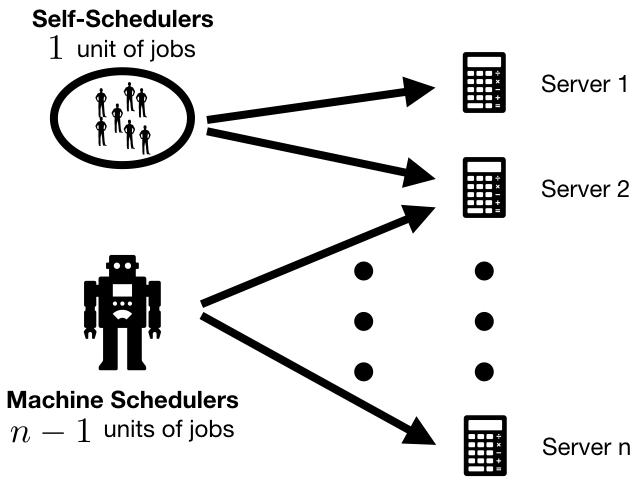}}
\caption{Depiction of constrained scheduling setup for Section~\ref{ssec:secLoss} and Theorems~\ref{thm:null} and~\ref{thm:intelligent}.
The $1$ unit of self-scheduled jobs can only access servers $1$ and $2$, while the $n-1$ units of machine-scheduled jobs can access any server numbered $2$ through $n$.}
\label{fig:schem}
\vspace{-5mm}
\end{figure}

Consider the setting of Theorem~\ref{thm:special}, in which there are $n\geq3$ identical servers with delay functions $\tau_i(x_i) = x_i$, but now suppose that the self-scheduled traffic can only access servers $1$ and $2$, and that each of the machine schedulers can access any server numbered $2$ through $n$ (that is, machine schedulers can access any server \emph{other} than $1$).
As before, let $r$ denote the mass of jobs controlled by the machine schedulers; to ensure that this system’s optimal operating point is identical to the system of Theorem~\ref{thm:special} for each $\alpha\geq0$, we enforce an additional constraint that $r=n-1$.
See Figure~\ref{fig:schem} for a depiction of this constrained system.

In this constrained context, our first question is this: if the machine schedulers apply the naive optimal scheduling policy informed by Theorem~\ref{thm:main}, how does this system's security compare to the unconstrained system of Theorem~\ref{thm:special}?
Theorem~\ref{thm:null} demonstrates that despite the fact that the machine schedulers control the vast majority of jobs, the new constraint completely nullifies the previous benefits of machine scheduling.

\begin{theorem} \label{thm:null}
For $n\geq3$, in the above system, let an attacker attack server $1$ with attack strength $\alpha\geq0$.
Let $\xt(\alpha)$ be a team equilibrium, and let $\xn(\alpha)$ be an uncoordinated self-scheduled equilibrium obtained by converting all jobs to selfish jobs.
Then we have\vs\vs
\begin{equation} \label{eq:null}
\C^{\alpha}(\xt(\alpha)) = \C^{\alpha}(\xn(\alpha)) = \min\left\{ \frac{n}{n-1}, 1+\frac{\alpha}{n} \right\}.
\end{equation}
\end{theorem}\vspace{2mm}
The proof of Theorem~\ref{thm:null} appears in the Appendix.\vspace{2mm}

That is, in the above system, even if the machine schedulers control a majority fraction of $(n-1)/n$ of the total traffic, simply being unable to access a single machine can dramatically increase the attacker’s ability to harm overall system performance.
In fact, for attacks with $\alpha\leq n/(n-1)$, it holds that no team equilibrium outperforms an unresponsive scheduling policy, or $\C^\alpha(\xt(\alpha))=\C^\alpha(\xb)$, just as in~\eqref{eq:baselineSelf}.

\begin{figure}
\centerline{\includegraphics[scale=0.5]{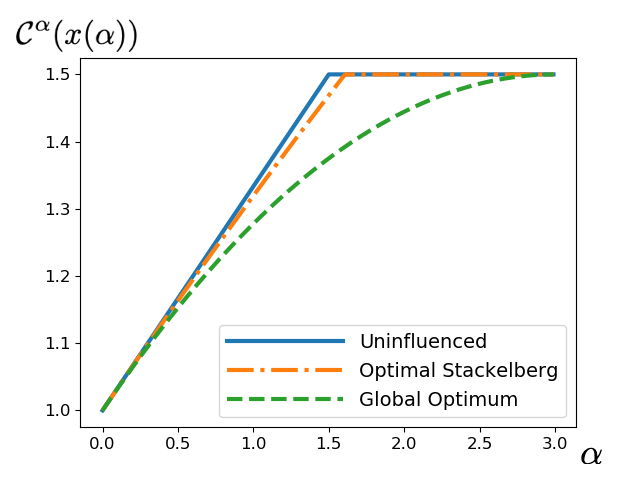}}
\caption{Normalized costs $\C^\alpha(x)$ for uninfluenced team equilibrium (Equation~\eqref{eq:null}, optimal Stackelberg equilibrium (Equation~\eqref{eq:xsC}), and globally-optimal scheduling profile (Equation~\eqref{eq:optC}) plotted with respect to $\alpha$ for $n=3$.
Note that for moderate values of $\alpha$, the Stackelberg equilibrium slightly outperforms the uninfluenced equilibrium, but only by a very small amount compared to the gains possible with global scheduling.}
\label{fig:stackEq}
\vspace{-5mm}
\end{figure}

To compare~\eqref{eq:null} with system optimal, consider Figure~\ref{fig:stackEq}; there, the solid (blue) trace corresponds to the constrained team scheduling in~\eqref{eq:null}, and the dashed (green) trace corresponds to system optimal scheduling.

Note that Theorem~\ref{thm:null} holds for the case that the machine schedulers behave in an uncoordinated manner, each attempting to optimize global performance given the choices of others.
What if, on the other hand, a machine scheduler had a ``bigger-picture’’ view of the situation, and could schedule its traffic \emph{in anticipation} of the self-schedulers’ response?

To understand the effect of such socially-aware machine scheduling, we model this as a Stackelberg equilibrium in which a single machine scheduler selects a policy for servers $\{2,\dots,n\}$ and then the selfish schedulers select between servers $1$ and $2$ on the basis of delay, reminiscent of~\cite{Bonifaci2010}.

Here, we denote the machine-scheduled policy by $x^{\rm m}=(x_1^{\rm m},\dots,x_n^{\rm m})\in (n-1)\Delta(n)$ and the selfishly-scheduled policy by $x^{\rm s} = (x_1^{\rm s},\dots,x_n^{\rm s})\in \Delta(n)$, and to model constrained scheduling, we enforce that $x_1^{\rm m}=x_3^{\rm s}=\cdots=x_n^{\rm s} = 0$.
For each $i$, we let $x_i=x_i^{\rm m}+x_i^{\rm s}$.%

The machine schedulers select a policy to minimize the cost function\vs
\begin{equation}
J(x^{\rm m},x^{\rm s}) = \C^{\alpha}\left(x^{\rm m}+x^{\rm s}\right).
\end{equation}
Given the machine’s choice of $x^{\rm m}$, the selfish schedulers choose between servers $1$ and $2$ to satisfy the Nash condition~\eqref{eq:selfishTeam} given in this case by
\begin{equation} \label{eq:nashStack}
\begin{aligned}
x_1^{\rm s} > 0 \ &\implies \ \tau_1(x^{\rm s}_1) + \alpha \leq \tau_2(x_2^{\rm s}+x^{\rm m}_2), \\
x_2^{\rm s} > 0 \ &\implies \ \tau_1(x^{\rm s}_1) + \alpha \geq \tau_2(x_2^{\rm s}+x^{\rm m}_2).
\end{aligned}
\end{equation}

Given $\alpha\geq0$, we say that a scheduling profile $\left(\hat{x}^{\rm m}(\alpha),\hat{x}^{\rm s}(\alpha)\right)$ is a \emph{Stackelberg equilibrium} if it satisfies
\begin{equation} \label{eq:stackCond}
\begin{aligned}
\hat{x}^{\rm m}(\alpha) 	&\in\argmin_{x^{\rm m}\in(n-1)\Delta(n)} J(x^{\rm m},x^{\rm s}(x^{\rm m})), \\
\hat{x}^{\rm s}(\alpha)  	&= x^{\rm s}(\hat{x}^{\rm m}).
\end{aligned}
\end{equation}
where $x^{\rm s}(x^{\rm m})$ denotes that the selfish jobs have scheduled according to~\eqref{eq:nashStack}, given the choice of $x^{\rm m}$.

%
%
%
The following theorem reports the optimal Stackelberg strategy, and demonstrates that only slight security gains are possible in this situation.

\begin{theorem} \label{thm:intelligent}
For $n\geq3$, in the above system, let an attacker attack server $1$ with attack strength $\alpha\geq0$.
The following scheduling policy enforces an optimal Stackelberg equilibrium satisfying~\eqref{eq:stackCond}:
\begin{equation}\label{eq:optStackPolicy}
x_2^{\rm m} = \left\{
\begin{array}{ll}
1-\frac{\alpha(n-2)}{2n}			& \mbox{if }\alpha<\frac{2n}{n-2}\left(2-\sqrt{\frac{2n}{n-1}}\right), \\
\frac{1}{n-1}	& \mbox{otherwise. }
\end{array}
\right.
\end{equation}
and for all $i>2$, $x_i^{\rm m}=(n-1-x_2^{\rm m})/(n-2).$
Let $\hat{x}(\alpha)$ be an optimal Stackelberg equilibrium enforced by policy~\eqref{eq:optStackPolicy}.
Then we have
\begin{equation}\label{eq:xsC}
\C^{\alpha}(\hat{x}(\alpha)) = \min\left\{ \frac{n}{n-1}, 1+\frac{\alpha}{n}-\frac{\alpha^2(n-2)}{8n^2} \right\}.
\end{equation}
\end{theorem}\vspace{2mm}
The proof of Theorem~\ref{thm:intelligent} appears in the Appendix.\vspace{2mm}

The optimal scheduling policy shown in~\eqref{eq:optStackPolicy} exhibits the following threshold behavior: when $\alpha$ is below the threshold, the optimal policy is to artificially increase the congestion on server $2$ in an effort to limit access by the selfish jobs.
When $\alpha$ increases above the threshold, the optimal policy discontinuously decreases the jobs scheduled on server $2$, since at that point the optimal Stackelberg equilibrium simply equals the team equilibrium from Theorem~\ref{thm:null}.
That is, when $\alpha$ is above the threshold, the costs of influencing behavior outweigh the benefits, and it becomes optimal to simply allow the self-routers to abandon the attacked server.

Clearly, the normalized cost in the Stackelberg case~\eqref{eq:xsC} is always less than the nominal uninfluenced cost from Theorem~\ref{thm:null}, but Figure~\ref{fig:stackEq} clearly illustrates that this improvement is quite small when compared to the optimal attainable value from~\eqref{eq:optC}.
That is, Theorem~\ref{thm:intelligent} suggests that the inefficiency introduced by scheduling constraints presents a fundamental impediment to system performance, and is thus deserving of further study.

\begin{figure}
\centerline{\includegraphics[scale=0.4]{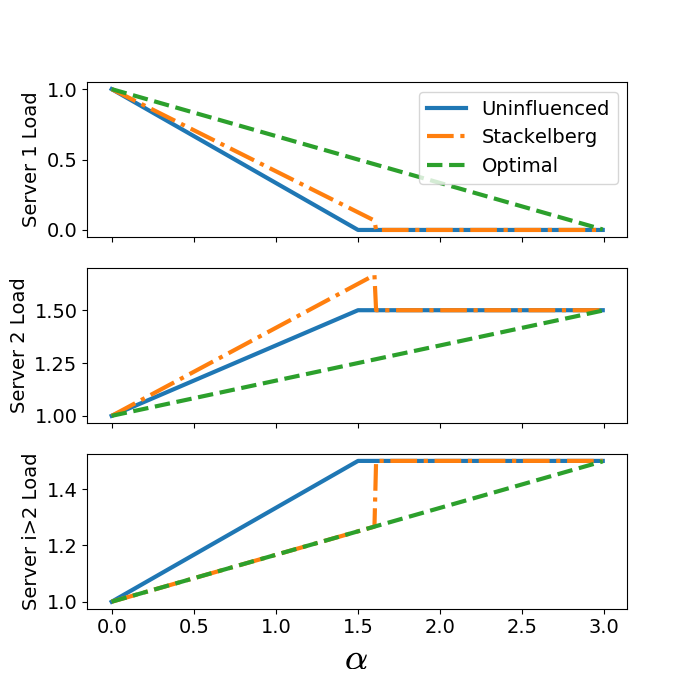}}
\caption{Server loads for each of the scheduling profiles discussed in Section~\ref{ssec:secLoss}, as a function of $\alpha$.
Note that the optimal (green dashed trace) reacts to $\alpha$ the most moderately of the three profiles, whereas in the Stackelberg case (orange dash-dot trace), the load on Server 2 increases rapidly in an attempt to limit the load of self-scheduled jobs on that server.}
\label{fig:stackEq2}
\vspace{-5mm}
\end{figure}

Furthermore, note that the optimal Stackelberg strategy can optimize for system performance only by deliberately degrading the performance of Server~$2$.
This is justifiable from the standpoint of overall system performance, since the inefficiencies caused by selfish scheduling are a direct result of too many selfish jobs selecting server~$2$.
However, the effect of this is that the selfish jobs experience a significantly-degraded quality of service for a wide range of $\alpha$.
This can be seen directly in the middle subplot of Figure~\ref{fig:stackEq2}; here, the dash-dot (orange) and solid (blue) traces capture both the load on Server~$2$ and the total cost experienced by selfish jobs in Stackelberg an uninfluenced settings, respectively.


\section{Conclusion}

The results in this paper represent an initial summary look at several possible issues that can arise when distributed autonomous decisionmaking architectures are integrated with those of self-interested human agents.
While certain systems are highly receptive to mixed human-autonomous decisionmaking (such as the unconstrained system of Theorems~\ref{thm:main} and~\ref{thm:special}), in other types of system, machine scheduling is essentially useless (in the sense that selfish scheduling would perform just as well, as in Theorem~\ref{thm:null}).
In such systems, even socially-aware machine schedulers (which make decisions in light of their impacts on other decisionmakers) need not improve the situation much.
As discussed following Theorem~\ref{thm:intelligent}, optimal socially-aware scheduling can even (perhaps paradoxically) cause significant increases in the costs experienced by the human system users.
Put together, this paper paints an optimistic picture of the possible benefits of human-machine teams in certain circumstances -- but it also tells a cautionary tale of some potential challenges and pitfalls of the teamed paradigm.

\bibliographystyle{ieeetr}
\bibliography{../library/library}

\appendix

\section*{Proof of Theorem~\ref{thm:main}}
The proof hinges on the fact that the problem of finding a team equilibrium can be reduced to finding a Nash flow for a heterogeneous price-sensitive user population in a parallel-network routing game under the influence of marginal-cost tolls.
Accordingly, we use techniques and appeal to several results from~\cite{Brown2017b} which pertain to this problem.
Given an instance of the team equilibrium problem, construct an instance of the Nash flow routing problem in the following way.
Model each server as a network link with delay function $\tau_i$.
Create two classes of users; one (corresponding to the selfishly scheduled jobs) with mass $n-r$ and cost function on server $i$ simply as $\tau_i(x_i)$.
The second class (corresponding to the machine-scheduled jobs) has mass $r$ and cost function on server $i$ given by $\tau_i(x_i)+x_i\tau’_i(x_i)$, where $\tau’_i$ denotes the first derivative of $\tau_i$.
That is, the machine-scheduled jobs are modeled by unit-price-sensitive users in a transportation network being charged marginal-cost tolls.

To see that the models are equivalent, simply note that given a choice of server by the selfish jobs, the optimization objective the machine schedulers can be modeled as the potential function of a potential game with server cost functions of $\tau_i(x_i)+x_i\tau’_i(x_i)$.
That is, if $\xt$ is a team equilibrium, the machine-scheduled jobs can be modeled as nonatomic agents playing a game with cost functions $\tau_i(x_i)+x_i\tau’_i(x_i)$.
Thus, any team equilibrium of the scheduling problem is a Nash flow for the corresponding routing game.

Given this, we appeal directly to Lemma~2.1 of~\cite{Brown2017b} which states (in the language of the present paper) that for any machine penetration level $r$ and attack level $\alpha\geq0$, any associated team equilibrium $xt(\alpha)$ satisfies\vs
\begin{equation}\vs
\C^{\alpha}(\xt(\alpha))\leq\C^\alpha(\xn(\alpha)), \nonumber
\end{equation}
proving the first inequality in~\eqref{eq:generalineq}.


To see that~\eqref{eq:globopt} is true, note the following.
Let $x^*$ be an optimal scheduling profile when $\alpha=0$.
Let $k$ be the index of a minimum-delay server under profile $x^*$, or \vs
\begin{equation}\vs
k\in \argmin_{j\in\{1,\dots,n\}} \tau_j(x_j^*).
\end{equation}
Observe that at this optimal profile, if all jobs currently scheduled on server $k$ are ``converted’’ to self-scheduled jobs, they will still (at least weakly) prefer server $k$, since by definition, for all $j\geq1$,\vs
\begin{equation}\vs
\tau_k(x_k^*) \leq \tau_j(x_j^*).
\end{equation}
That is, setting $\bar{r}=\sum_{i\neq k}x_i^*$, there exists an optimal team equilibrium $\xt$ with $r=\bar{r}<n$ since the $n-\bar{r}$ selfish jobs can be considered to be selecting server $k$.
Note that even if $k=1$, it must be true that $x^*_1\leq \bar{r}$.
For the remainder of the proof, any team equilibrium is assumed to be for a population in which $r=\bar{r}$.

As discussed in the proof of Lemma~5.2 in~\cite{Brown2017b}, for this population it holds that for any nontrivial attack with $\alpha>0$ that in any team equilibrium, selfish schedulers will select server $1$ (the attacked server) if and only if no machine schedulers select any other server, or $\xt_1(\alpha)>\bar{r}$.
However, since all delay functions (and their marginal costs) are nondecreasing, any team equilibrium with $\alpha>0$ must have $\xt_1(\alpha)<\xt_1(0)$ and $\xt_i(\alpha)\geq\xt_i(0)$ for all $i>1$.
Due to these monotonicity properties, since $\xt(0)=x^*$, we thus have that for all $\alpha>0$, $\xt_1(\alpha)\leq\bar{r}$, meaning that no selfish schedulers select server $1$ in any team equilibrium.

To complete the proof of~\eqref{eq:globopt}, we show that when $\alpha>0$, in every $\xt(\alpha)$ it must be the case that every server has a positive mass of machine-scheduled jobs.
This is because the mass of self-scheduled jobs is $n-\min_i\{x_i(0)\}$, and for all $i>1$, $\alpha>0$ implies that $\xt_i(\alpha)>\xt_i(0)$.
Recalling again the previously-referenced statements from~\cite{Brown2017b}, if all servers have a positive mass of machine-scheduled jobs, the team equilibrium must satisfy~\eqref{eq:globopt}.
%
%
%
%
\hfill\QED

\section*{Proof of Theorem~\ref{thm:special}}

First we show that for any machine penetration level $r$ and attack level $\alpha$, every associated team equilibrium $\xt(\alpha)$ has\vs 
\begin{equation}\vs
\mbox{$\xt_i=\xt_j$ for all $i,j>1$.} \label{eq:equals}
\end{equation}
Suppose that $x$ is an aggregate scheduling profile in which $x_i>x_j$ for some $i,j>1$.
Note that if any selfish jobs are scheduled on server $i$, then~\eqref{eq:selfishTeam} implies that $x$ is not a team equilibrium, since there exists a server with lower cost (namely, server~$j$): $\tau_j(x_j)<\tau_i(x_i)$.
Further, note that $\tau_j(x_j)<\tau_i(x_i)$ does not optimize the machine objective of~\eqref{eq:optTeam} since a small mass of jobs transferred from $i$ to $j$ would decrease the system's total cost.
That is, if any machine has scheduled on server $i$, it follows that $x$ is not a team equilibrium.
Thus, for $i,j>1$, every team equilibrium has  $\xt_i=\xt_j$.

For any $\alpha\geq0$, it is straightforward to verify that the globally optimal scheduling policy $x^*$ has $x^*_1 = \max\{0,1-\frac{\alpha(n-1)}{2n}\}$, and the remaining jobs divided evenly among servers~$2$ through~$n$, so that for $i>1$, $x_i^*=\min\{n/(n-1),1+\frac{\alpha}{2n}\}$.
Let the machine penetration level meet or exceed $x_1^*$: $r\geq1-\frac{\alpha(n-1)}{2n}$.
Given this $r$, let $x$ be an aggregate profile with $x_1<x_1^*$ and the remaining jobs divided equally among servers~$2$ through~$n$.
In $x$, some machine must be scheduling a positive mass of jobs on a server other than $1$; this machine could improve the value of its objective by shifting some jobs to server $1$ to bring $x_1$ closer to $x_1^*$ -- implying that $x$ is not a team equilibrium.
On the other hand, let $x'$ be an aggregate profile with $x'_1>x_1^*$ and the remaining jobs divided equally among servers~$2$ through~$n$.
In $x'$, it is straightforward to verify that $\tau_1(x'_1)+\alpha>\tau_2(x'_2)$, implying that if selfish jobs are selecting $1$, $x'$ cannot be a team equilibrium.
Further, if all of the jobs on server $1$ are machine-scheduled, some of them could be shifted to servers $2$ through $n$ to bring the total cost of the profile closer to optimal.
Thus, $x'$ is not a team equilibrium, and it must be the case that when $r\geq1-\frac{\alpha(n-1)}{2n}$, every team equilibrium is globally-optimal;~\eqref{eq:optC} may be verified by substituting the aforementioned values of $x_1^*$ and $x_i^*$ into~\eqref{eq:attackCost}.

The proofs of the remaining statements in Theorem~\ref{thm:special} follow similar logic.
For~\eqref{eq:strict}, there are not enough machine schedulers to ensure that $\xt_1=x_1^*$, but there are enough to ensure that no selfish jobs schedule on server $1$.
Thus, in this intermediate case, every team equilibrium has $\xt_1\in\{0,r\}$ and $\xt_i=(n-\xt_1)/(n-1)$, and~\eqref{eq:strict} follows algebraically.

Finally, for~\eqref{eq:selfish}, there are so few machine-scheduled jobs in this case that the selfish jobs are all indifferent between all links and ensure that $\tau_1(\xt_1)+\alpha=\tau_i(\xt_i)$ for all $i\geq 1$, yielding $\xt_1=\max\{1-\alpha(n-1)/n,0\}$ and~\eqref{eq:selfish}.
\hfill\QED

\section*{Proof of Theorem~\ref{thm:null}}

Here, we may repeat the argument from the proof of Theorem~\ref{thm:special} that any team equilibrium in this case has \vs\vs
\begin{equation}\vs
\mbox{$\xt_i=\xt_j$ for all $i,j>1$.} \label{eq:equals2}
\end{equation}
The key feature of Theorem~\ref{thm:special} that allowed a small mass of machine-scheduled jobs to enforce optimality was that the machine-scheduled jobs could access the attacked server and effectively displace the selfish jobs.
In the present constrained setting, this is no longer possible since machine schedulers cannot access server~$1$.

Thus, the equilibrium condition of the selfish jobs guarantees that for low $\alpha$, in any team equilibrium $\xt(\alpha)$, $\xt_1+\alpha = \xt_2$.
Combining this with~\eqref{eq:equals2} ensures that $\xt(\alpha)$ is identical to a fully-selfish equilibrium $\xn(\alpha)$, and~\eqref{eq:null} may be computed as in the third part of the proof of Theorem~\ref{thm:special}.
\hfill\QED

\section*{Proof of Theorem~\ref{thm:intelligent}}

\begin{lemma} \label{lem:StackConstr}
For every $\alpha\geq0$, there exists an optimal Stackelberg equilibrium in which \vs
\begin{equation} \label{eq:optConstraint}\vs
x_1+\alpha\geq x_2.
\end{equation}
\end{lemma}\vspace{2mm}

\begin{proof}
This is immediately true for $\alpha=0$, when the unique optimal Stackelberg equilibrium has $\hat{x}_i=1$ for all $i$.
Therefore, let $\alpha>0$ and let $x$ be any scheduling profile such that $x_1+\alpha< x_2$.
If $x$ is a Stackelberg equilibrium, the selfish schedulers’ incentive constraint in~\eqref{eq:stackCond} implies that ${x}_2=0$, or that ${x}_1=1$.
From here, it is simple to show that the normalized cost of this is lower-bounded by\vs
\begin{equation}\vs
\C^\alpha(x) \geq 1+\frac{\alpha}{n},
\end{equation}
satisfied with equality when $x_i=1$ for all $i=1$.

We will show a machine-scheduled policy $\tilde{x}^{\rm m}$ which weakly outperforms profile $x$ and always has $\tilde{x}^{\rm m}_1+\alpha\geq \tilde{x}^{\rm m}_2$, implying that the optimal Stackelberg equilibrium can always be selected to satisfy $x_1+\alpha\geq x_2$.
Let $\tilde{x}^{\rm m}_2=1/(n-1)$, and let $\tilde{x}^{\rm m}_i=n/(n-1)$ for all $i>2$.
Clearly, for all $\alpha$, $\tilde{x}_1^{\rm s}=1$ cannot constitute a Nash response by the self-scheduled jobs, as this would yield $\tilde{x}_1+\alpha>\tilde{x}_2$, and selfish jobs would have a strong incentive to deviate to server $2$.
Thus, it must be the case that $\tilde{x}_2^{\rm s}>0$, which by the Nash condition in~\eqref{eq:nashStack} implies that $\tilde{x}_1+\alpha \geq \tilde{x}_2$.
Further manipulations yield the fact that $\C^\alpha(\tilde{x}) \leq 1+{\alpha}/{n}.$
Clearly, any optimal Stackelberg equilibrium $\hat{x}$ must weakly outperform $\tilde{x}$, implying that for every~$\alpha\geq0$, there exists an optimal Stackelberg equilibrium~$\hat{x}$ satisfying~\eqref{eq:optConstraint}.
\end{proof}

\subsection*{Proof of Theorem~\ref{thm:intelligent}.}


An optimal Stackelberg equilibrium $\hat{x}$ is a solution of the following optimization problem:\vs
\begin{equation} \label{opt:stack}
\begin{aligned}
&\underset{x}{\mbox{minimize}}	&& f(x) = \sum_{i=1}^nx_i^2 + \alpha x_1, \\
&\mbox{subject to } 				&& \sum_{i=1}^nx_i = n, \\
&							&& x_1 + \alpha \geq x_2, \\
& 							&& x_1\geq 0.
\end{aligned}
\end{equation}

The first constraint represents flow conservation; the second constraint is a consequence of Lemma~\ref{lem:StackConstr}.
It can easily be shown that the optimum must have $x_i>0$ for all $i>1$, leaving only the relevant nonnegativity constraint of $x_1\geq0$.
Associating KKT multipliers $\lambda\geq0$, $\mu_1\geq0$ and $\mu_2\geq0$ with the three constraints in~\eqref{opt:stack}, we have
 first-order optimality conditions given by\vs
\begin{align}
2x_1 + \alpha - \lambda + \mu_1 +\mu_2  	&=0, \label{eq:link1} \\
2x_2 - \lambda - \mu_1 			&= 0, \label{eq:link2} \\
 2x_i - \lambda 						&=0,\ \forall i>2, \label{eq:linki} \\
 \mu_1(x_2-x_1-\alpha) 			&=0, \label{eq:slack1} \\
 \mu_2x_1 						&= 0. \label{eq:slack2} 
\end{align}
Note the following:
\begin{enumerate}
\item $\mu_1>0$ implies that $x_1+\alpha = x_2$, which (due to the selfish jobs’ Nash conditions) implies that $x_1\geq0$, so we may consider $\mu_2=0$.
\item $x_1+\alpha > x_2$ implies that $x_1=0$, in which case $\mu_2\geq0$.
\end{enumerate}
That is, without loss of generality we may assume that $\mu_1>0 \iff \mu_2=0$.
With that in hand, when $\mu_1>0$ (and $\mu_2=0$), the local optimizer $x^*(\alpha)$ (for $\alpha\leq4n/(3n-2)$) satisfies\vs\vs
\begin{align}
x^*_1 	&= 1-\frac{\alpha(3n-2)}{4n}, \\
x^*_2 	&= 1+\frac{\alpha(n+2)}{4n}, \\
x^*_i 		&= 1+\frac{\alpha}{2n}, \ \forall i>2,
\end{align}
and the corresponding normalized cost can be derived as
\begin{equation}
\C^{\alpha} (x^*(\alpha)) = 1+\frac{\alpha}{n}-\frac{\alpha^2(n-2)}{8n^2}.
\end{equation}

When $\mu_1=0$ and $\mu_2>0$, the resulting equilibrium resolves to the under-influenced team equilibrium $\xt(\alpha)$ of Theorem~\ref{thm:null}, with normalized cost satisfying that of~\eqref{eq:null}.

Thus, the optimal Stackelberg equilibrium $\hat{x}(\alpha)$ has normalized cost equal to that in~\eqref{eq:xsC}.
Finally, equations~\eqref{eq:link1}-\eqref{eq:linki}, combined with the selfish jobs’ conservation constraint, can be manipulated to show~\eqref{eq:optStackPolicy}.
\hfill \QED


\end{document}

%% file: philipMacros.tex
\newcommand{\xb}{x^{\rm b}}
\newcommand{\xt}{\bar{x}}
\newcommand{\xn}{x^{\rm n}}
\newcommand{\C}{\mathcal{C}}

\DeclareMathOperator*{\arginf}{arg\,inf}

\DeclareMathOperator*{\argmin}{arg\,min}

\newcommand{\vs}{\vspace{-1mm}}

